\documentclass[12pt]{iopart}

\usepackage{graphicx}

\usepackage{iopams}  
\expandafter\let\csname equation*\endcsname\relax
\expandafter\let\csname endequation*\endcsname\relax
\usepackage{amsmath}
\usepackage{braket}
\usepackage{bm}

\newcommand{\nred}{\textcolor{black}}

\usepackage{subcaption}
\usepackage{pgfplots}
\usepackage{siunitx,cite}
\usepackage{xcolor}
\usepackage[breaklinks=true,colorlinks=true,linkcolor=blue,urlcolor=blue,citecolor=blue]{hyperref}
\usepackage{float}
\usepackage{filecontents}
\usepackage{
	pst-plot,
	pst-ode
}

\begin{document}

\title[Quantum search with a CTQW in momentum space]{Quantum search with a continuous-time quantum walk in momentum space}

\author{Michele Delvecchio$^1$, Caspar Groiseau$^{2,3}$, Francesco Petiziol$^{1,4}$, Gil S Summy$^{5,6}$, and Sandro Wimberger$^{1,4}$}
\address{
     $^1$ Dipartimento di Scienze Matematiche, Fisiche ed Informatiche, Universit\`a di Parma, Parco Area delle Scienze 7/A, 43124 Parma, Italy\\
	$^2$ Department of Physics, University of Auckland, Private Bag 92019, Auckland, New Zealand \\
	$^3$ Dodd-Walls Centre for Photonic and Quantum Technologies, New Zealand \\
	$^4$ INFN, Sezione di Milano Bicocca, Gruppo Collegato di Parma, 43124 Parma, Italy \\
	$^5$ Department of Physics, Oklahoma State University, Stillwater, Oklahoma 74078-3072, USA \\
	$^6$ Airy3D, 5445 Avenue de Gasp\'e Suite 230, Montr\'eal, Qu\'ebec H2T 3B2, Canada
}
\ead{sandromarcel.wimberger@unipr.it}
\vspace{10pt}
\begin{indented}
\item[\today]
\end{indented}

\begin{abstract}
The atom-optics kicked rotor can be used to prepare specific momentum distributions on a discrete basis set. We implement a continuous-time quantum walk and a quantum search protocol in this momentum basis. In particular we propose ways to identify a specific marked state from the final momentum distribution after the walker's evolution. Our protocol is guided by current experimental possibilities making it accessible to experimentally implemented quantum walks with Bose-Einstein condensates. 

\end{abstract}

\section{Introduction}
\label{introduction}

For over 25 years quantum algorithms \cite{QA2016} have played a central role in the quest to speed up computational power, due to their higher performance with respect to classical analogues. Quantum search is one of the most studied algorithms \cite{QA2016, grover1997quantum, Portugal}. It may be implemented using a quantum walk (QW), see e.g. Refs. \cite{childs2004spatial, shenvi2003quantum, Portugal, QA2016}. Following this direction, in this paper we propose  a quantum search protocol suitable for the experimental realization of the quantum kicked rotor (QKR), also known as atom-optics kicked rotor (AOKR), see e.g. \cite{Raizen1999, sadgrove2011pseudoclassical}. In Ref. \cite{equivalence} some of us observed that the QKR at resonance conditions realizes a simple one-dimensional continuous-time quantum walk (CTQW). \nred{In the jargon of QWs, a CTQW is characterized by the absence of the degree of freedom given by the coin, as described, e.g., by \cite{Portugal, farhi1998quantum}. This is in contrast to a so-called discrete-time QW in which an additional coin degree of freedom is periodically flipped \cite{Portugal}. We deliberately keep this nomenclature in order to connect to the community of QWs. This is also justifiable by the fact that the physical realisations of periodically driven Floquet problems \cite{Floquet1965} are nevertheless evolving continuously in time, see e.g. our Hamiltonian of the QKR below in Eq. \eqref{eq:ham-qkr}.} We now take advantage of the versatility of the AOKR experiments in order to implement a quantum search protocol based on such a CTQW.

While our main goal is not to improve the performance of highly specialist computer science algorithms, but rather to present a method by which standard experiments with a Bose-Einstein condensate (BEC) may be used to perform a simple quantum search, within the basis set of a one-dimensional grid of discrete momentum states formed from a BEC. This basis is defined by the periodic potential kicking the BEC and changing its momenta by discrete steps in units of two-photon recoils.

The paper is organised as follows: in Sec. \ref{sec:2}, after a brief introduction of the temporal evolution of the AOKR at quantum resonance conditions, we present the search protocol adapted to our experimental system. Consequently, we propose two different techniques to obtain the desired state from measurements of the momentum distributions after specific evolutions. In sec. \ref{sec:3}, we discuss the temporal scaling of our search protocol, and we finish with general remarks on recurrence and hitting times of our realisation of a CTQW used for searching. Sec. \ref{sec:concl} concludes the paper.

\section{Implementation of a quantum search using the AOKR}
\label{sec:2}

\subsection{AOKR evolution}
\label{subsec:2:1}

The Hamiltonian describing the dynamics of the quantum kicked rotor \cite{Casati, fishmanDL, wimberger} is usually represented in dimensionless units as
\begin{eqnarray}
\label{eq:ham-qkr}
\hat{\mathcal H} = \frac{\hat p^2}{2}+k\cos \hat \theta \sum_{j=1}^{T}\delta(t-j\tau) \, ,
\end{eqnarray}
where $p$ is the (angular) momentum, $\theta$ the position (angle), $k$ the kick strength, and $\tau$ the period of the kicks.
As discussed in \cite{equivalence}, we work with the QKR at resonance conditions. In particular, we will consider the principal quantum resonances (QR) of the kicked rotor \cite{Izr1990}, and for simplicity we set $\tau = 4 \pi$ in our units. The experimental realisation of the AOKR uses cold atoms or Bose-Einstein condensates moving in position space which are periodically kicked by the application of an optical lattice \cite{sadgrove2011pseudoclassical}. Then the dynamics is still described by the Hamiltonian \eqref{eq:ham-qkr} as long as the quasimomenta which are initially populated are zero or close to zero, see the experimental literature for details \cite{Summy2006, gil2008, WRH2013, Ratchet2017, PRL2018, PRA2019}.

At QR, the evolution is only given by the time-periodic kicks represented by the one-cycle Floquet operator
\begin{eqnarray}
\label{eq:kick}
	 \hat{\mathcal U} = e^{-i k \cos \hat \theta} \, ,
\end{eqnarray}
where $\hat{\theta}$ represents the angular (spatial) coordinate of the system. The evolution occurs in (angular) momentum space at discrete integers represented by the (angular) momentum operator $\hat p = \hat n =-\mathrm{id/d}\theta$, with periodic boundary conditions. At QR, the momentum distribution of a single momentum initial state, e.g. $n_0=0$, expands symmetrically around its initial value and displays ballistic expansion, i.e., with a standard deviation proportional to the number of applied kicks \cite{Izr1990, sadgrove2011pseudoclassical}. In Ref. \cite{equivalence}, it was proven that such momentum distributions are identical to the walker's distribution after a one-dimensional CTQW of the same duration (measured in discrete numbers of kicks). Such walks occur in the \textit{forward evolution} and typical plots of the corresponding momentum distributions are found in Refs. \cite{sadgrove2011pseudoclassical, Weiss2015}.

QR gives the experimental possibility of performing a \textit{backward evolution}, applying the operator $\hat{\mathcal{U}}^\dagger$ after the forward one has been completed. The representation of this forward-backward evolution is shown in Fig. \ref{fig:1a}. The inversion of the motion has been realised experimentally in Refs. \cite{PRL2018, PRA2019}, building on the fact that the effective sign in Eq. \eqref{eq:kick} can be changed by shifting the potential by $\pi$. As we will see in the next subsection, this control of the dynamics is the key idea of our search protocol.

\subsection{Preparation of the basis for the search}
\label{subsec:2_2}

The first step of our protocol is a preparation step. Here, the basis must be populated within a certain window in momentum space. An AOKR experiment typically starts out with atoms (the BEC) at rest with momentum zero. This initial state can be broadened simply by time evolution under QR, using the successive application of the Floquet operator \eqref{eq:kick}.  

\begin{figure}[th]
	\centering
	\begin{subfigure}[]{.47\linewidth}
		\includegraphics[width=\linewidth]{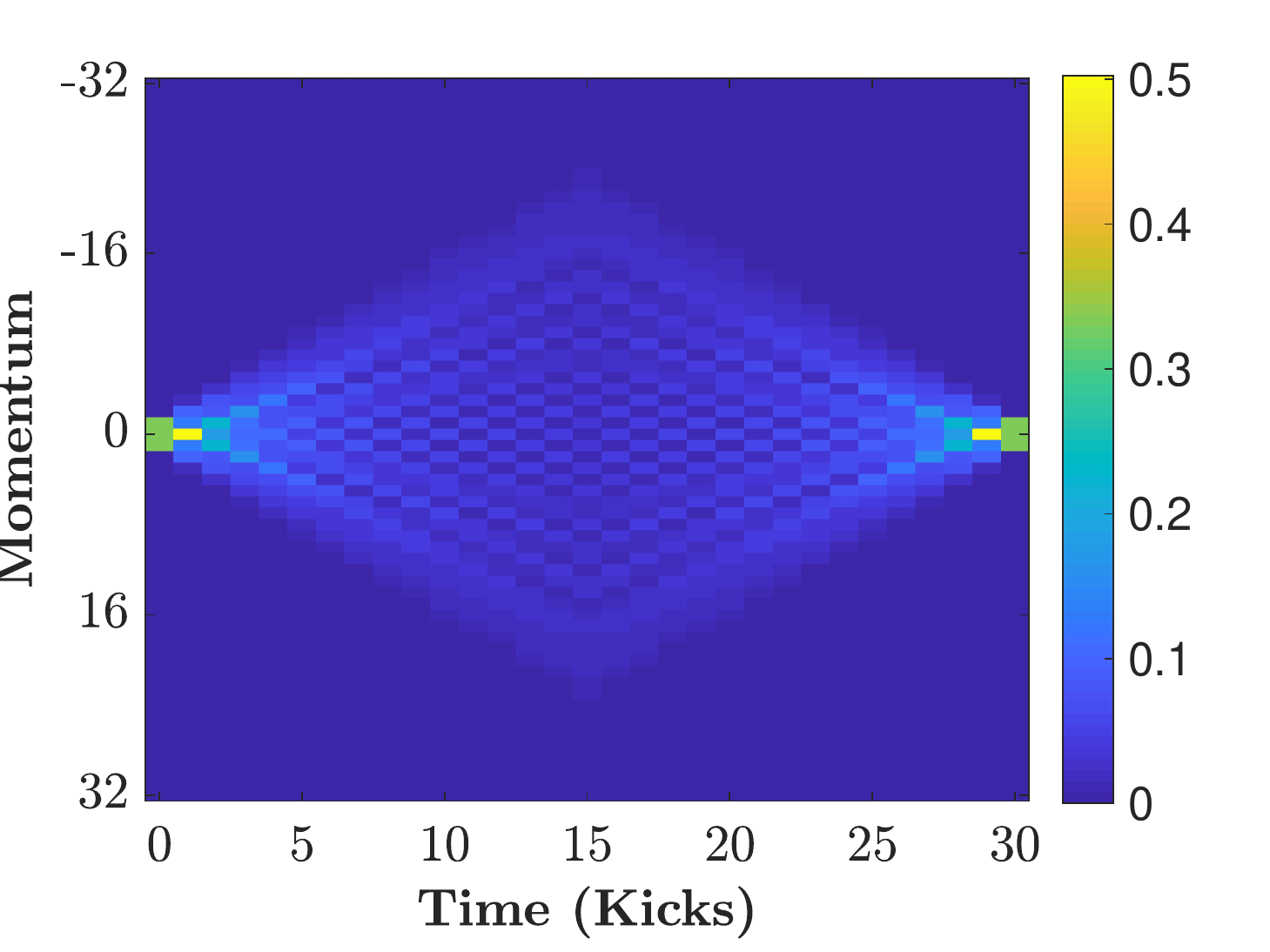}
		\caption{}
		\label{fig:1a}
		\end{subfigure}
	\begin{subfigure}[]{.47\linewidth}
		\includegraphics[width=\linewidth]{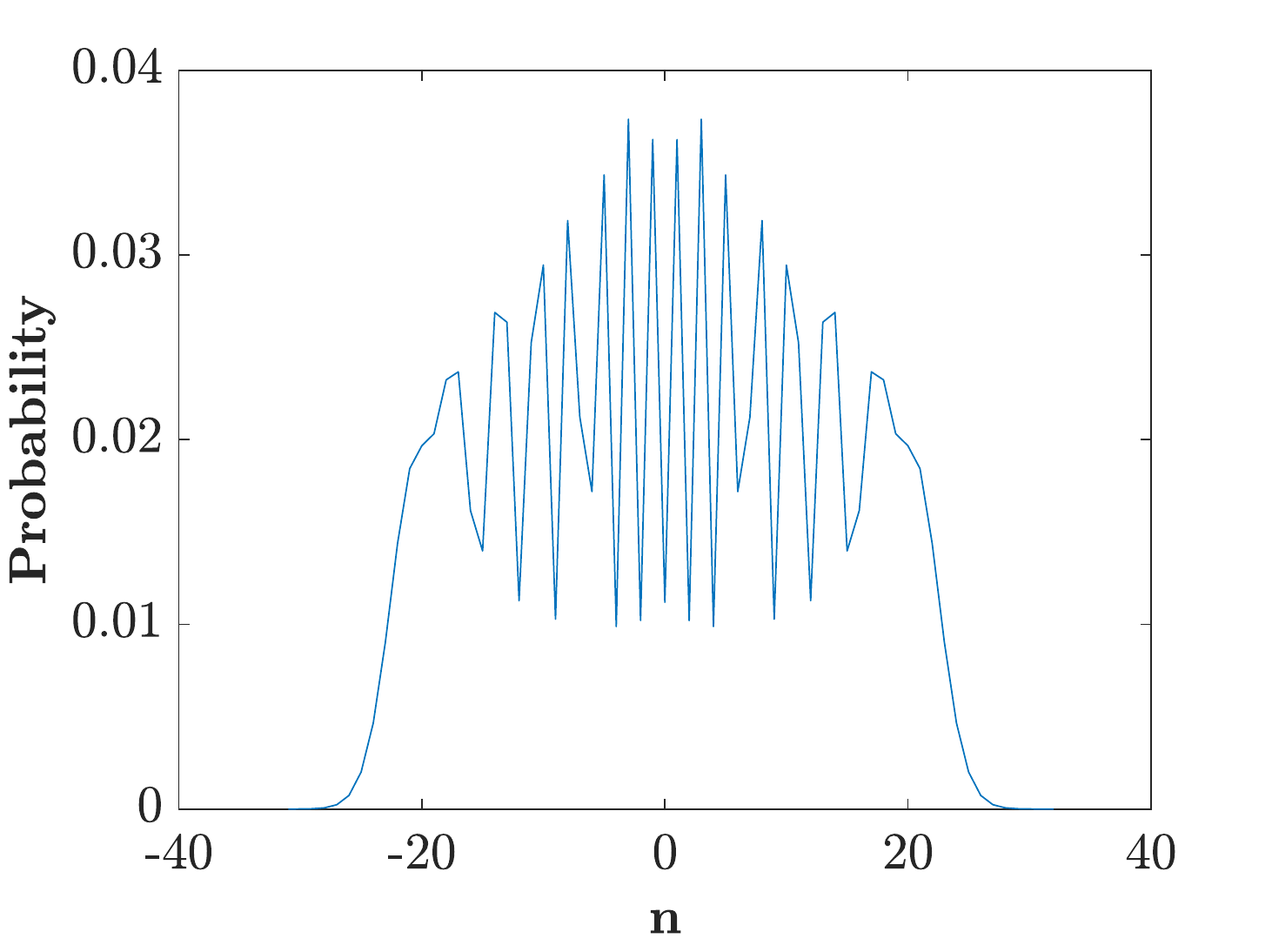}
		\caption{}
		\label{fig:1b}
		\end{subfigure}
       \begin{subfigure}[]{.47\linewidth}
		\includegraphics[width=\linewidth]{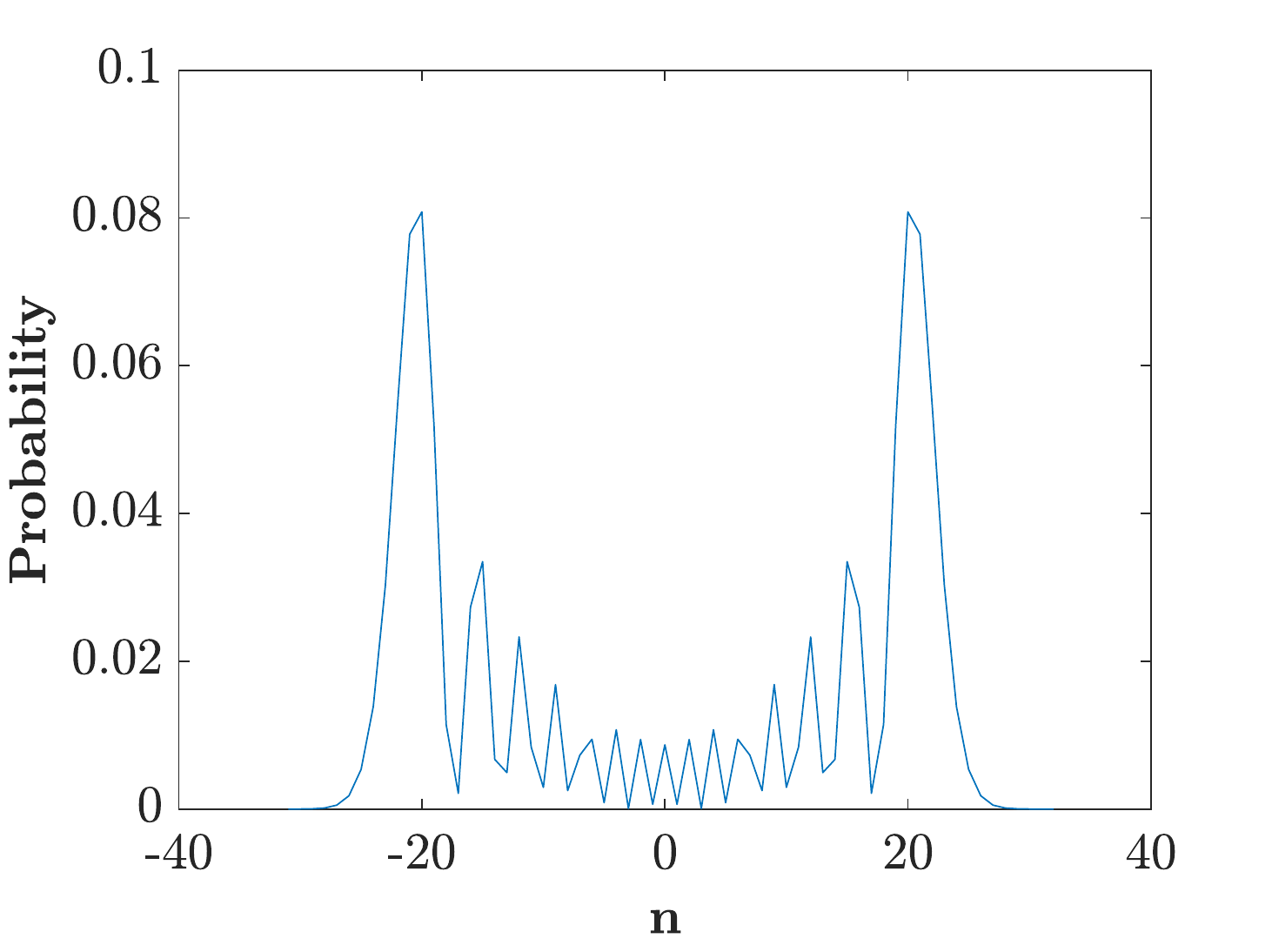}
		\caption{}
		\label{fig:1c}
		\end{subfigure}
	\begin{subfigure}[]{.47\linewidth}
		\includegraphics[width=\linewidth]{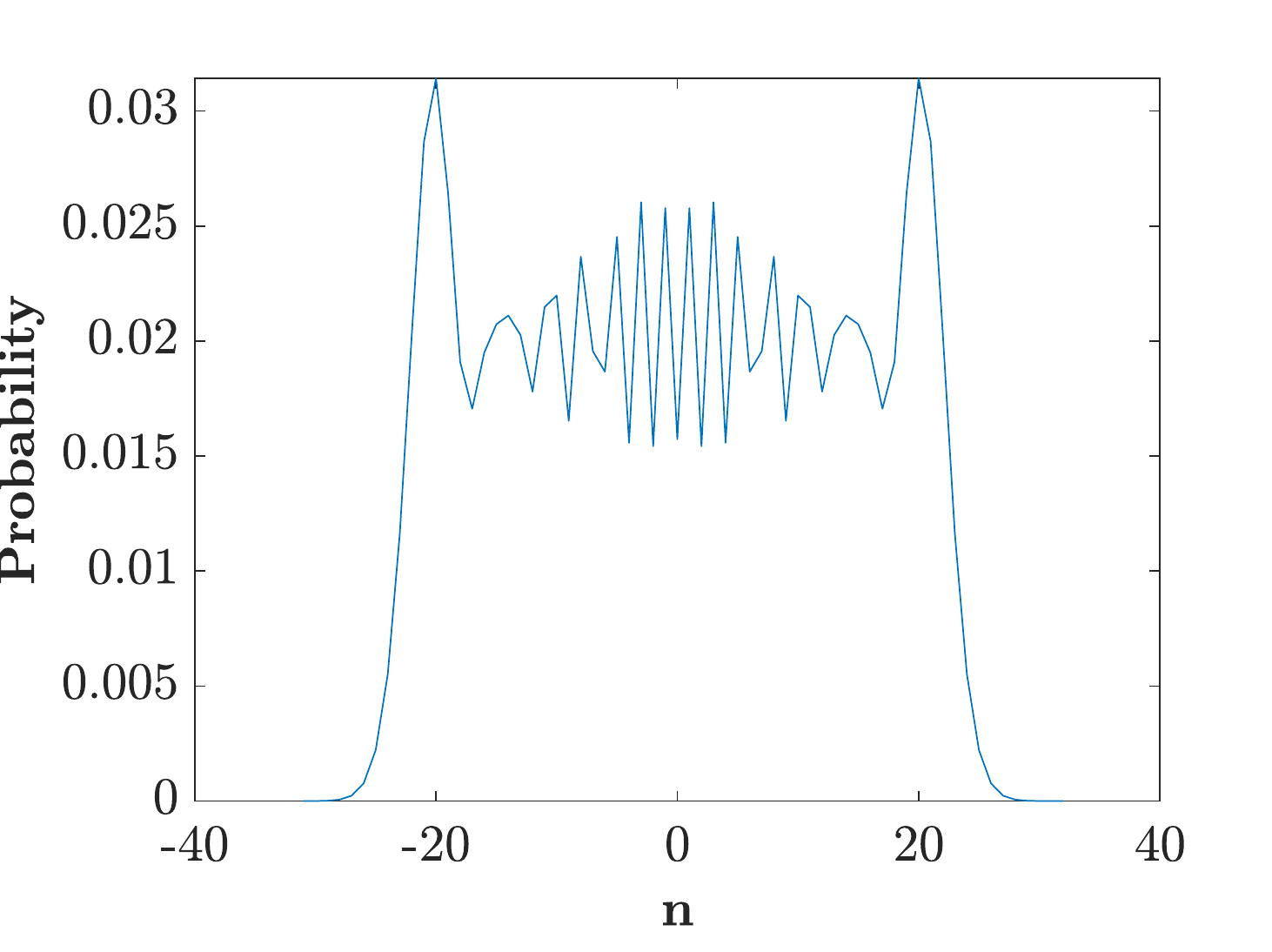}
		\caption{}
		\label{fig:1d}
		\end{subfigure}
		\caption{(a) Forward-backward evolution of an initial state composed of three momentum states taken around zero. Under ideal QR conditions, the final state at $t=30$ reproduces the initial state perfectly. The time inversion took place at $t=15$. (b,c,d) Momentum distributions after application of 15 kicks for initial states composed of the three momentum states $n=-1, 0, 1$, but with different coefficients: $c_{-1}=c_0=c_1=1/\sqrt{3}$ (b), $c_{1}= c_0= - c_{-1} = 1/\sqrt{3}$ (c), and $c_{-1}=0.4815, c_0=0.7323, c_1=0.4815$ (d). The total widths of the cases (b,c,d) are equal, depending only on the kicking strength $k$ and the number of kicks $t=15$, whilst the form of the distributions change. For our search algorithm, we will use the distribution (b) for its flatness at the center and its larger weights in the central part. For searches in a broader window, the parameters $k$ and $t$ can be increased, see main text.
		}
		\label{fig:1}
\end{figure}

The reason for this first step is that, in search algorithms, the initial state typically is a uniform "flat" superposition of the basis states, see e.g. Portugal's book \cite{Portugal}. With our system, we can reach this situation starting from a narrow distribution (superposition of one to several eigenstates) and optimizing the linear combination in order to obtain a distribution as flat as possible after a fixed number of kicks. \nred{For the purpose, the optimization consists in minimizing a cost function which quantifies the deviation of the distribution of the QW from the uniform distribution. This is defined as $\mathcal{E}(\bm{C})=\sum_{n=-N/2}^{N/2}\left\lvert P(n,\bar{t};\bm{C})-u_N\right\lvert$ where $N$ is the number of basis states considered, $u_N$ is the uniform distribution, $\bar t$ is the number of kicks, and $\bm{C}$ the vector of initial coefficients over which we run the minimization. This procedure is discussed in detail in Ref. \cite{equivalence}.} Figures \ref{fig:1}(b,c,d) show data for several momentum distributions obtained by the initial states of width three. While in the symmetric case of a QW the typical distribution is peaked at the flanks, changing the coefficients of the three initial states induces a temporal interference pattern which can lead to nearly flat distributions. The width of the latter distribution is directly proportional to the kicking strength $k$ and the preparation time \cite{WGF2003, sadgrove2011pseudoclassical}, i.e. the number of steps in the CTQW. This explains the aforementioned proportionality between the widths of the distribution and the kick number. For our search algorithm we may either use the distribution from Fig. \ref{fig:1b} or Fig. \ref{fig:1d} because of their flatness in the center, i.e. between momenta of $n \approx -10$ to $n \approx 10$. We will use in the following \nred{the one shown in (b) for simplicity}. For searches in a broader window, the parameters $k$ and $t$ can be increased accordingly.

\subsection{Search protocol}
\label{subsec:2_3}

\begin{figure}[th]
	\centering
		\begin{subfigure}[]{.47\linewidth}
		\includegraphics[width=\linewidth]{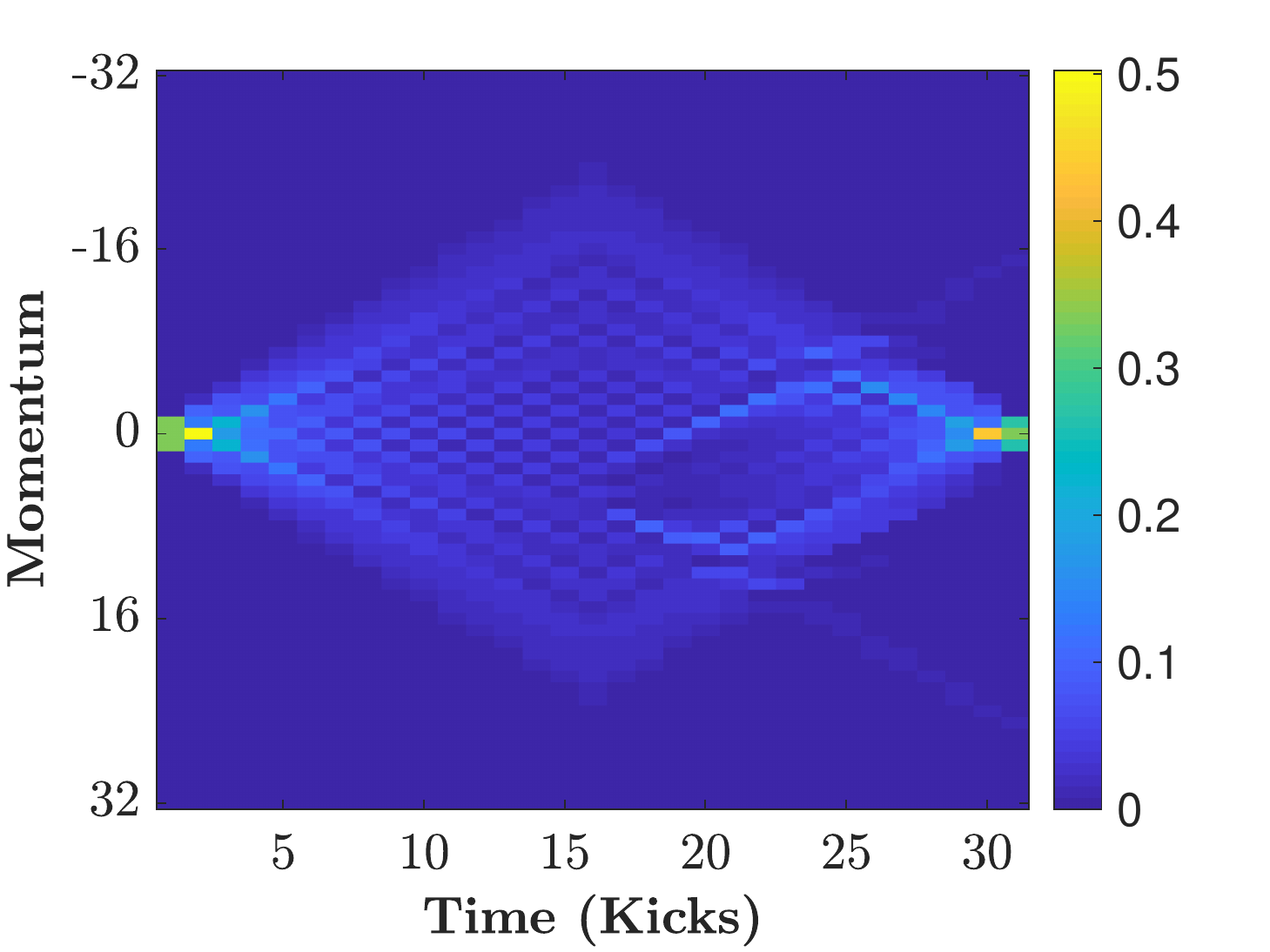}
		\caption{}
		\label{fig:2a}
		\end{subfigure}
	\begin{subfigure}[]{.47\linewidth}
		\includegraphics[width=\linewidth]{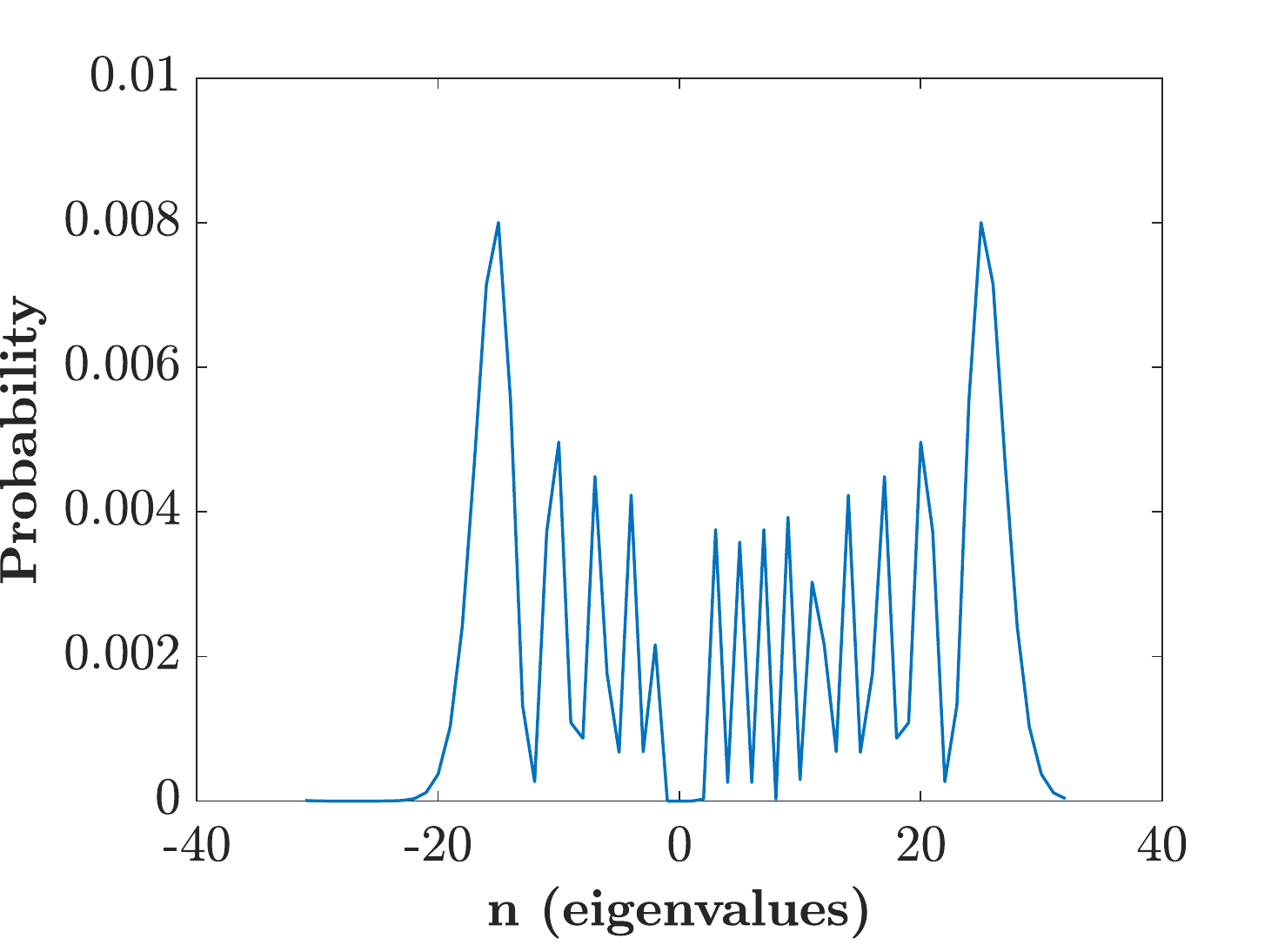}
		\caption{}
		\label{fig:2b}
	\end{subfigure}
		\caption{(a) Altered forward-backward evolution, with respect to the previous figure, now marking a specific state (with $n=5$ in this case) at $t=15$ before inverting the evolution. Now the final distribution at $t=30$ is a result of the walks performed by the marked state (forward) as well as the initial state (backward). (b) Momentum distribution at $t=30$ after suppressing the states around zero momenta, which would have much larger contributions.}
		\label{fig:2}
\end{figure}

The general idea of our protocol is as follows: we prepare the initial state performing the forward evolution, as explained in the previous subsection; then we mark a generic state by rotating its phase by $\pi$ \cite{Sadgrove2012}, i.e. by multiplying its coefficient by the factor $-1$, respective to all other states; next we perform the backward evolution, and finally we extract the marked state from the final distribution. The latter step can be done in different ways, as we will discuss below. Figure \ref{fig:2}(a) shows the backward evolution after the marking of a specific target state at time $t=15$. After that we need to extract the information obtained from the final momentum distribution, at $t=30$ in Fig. \ref{fig:2}(a), in order to find the target state. This latter step is explained in the following subsection.

\subsection{Extracting the marked state}
\label{subsec:2_4}

As we can notice from the evolution in Fig. \ref{fig:2a}, the marked state creates a new CTQW propagating forward, together with the original backward evolution. Both of these walks, the spreading one and the refocusing one, interfere to form the distributions observed in Fig. \ref{fig:2a} for different times $t=16\ldots30$. At $t=30$, however, most of the probability is concentrated around the originally populated states near zero momentum. Since we aim to study the walk generated by the marked state, we need to suppress the contribution of the initial state in the interference pattern. To do so, we \nred{may} adopt two strategies: 
\nred{
\begin{enumerate}
\item[\em (i)] Velocity/momentum-selective Raman transitions \cite{Raman} can be used to get rid of a certain window of momentum states by "kicking" those states toward much higher 
momenta and hence outside the experimental observation window. 
\item[\em (ii)] We also can simply subtract the unperturbed distribution (where no state marking occurred) from Fig. \ref{fig:1a} from the one seen in \ref{fig:2a} for any time $t = 16 \ldots 30$. At least in our theoretical calculation or numerical evolutions such a way of disregarding the contribution of the initial walk is always possible.
\end{enumerate}}

\noindent\nred{So now we can detect the marked state following two approaches:}

\begin{figure}[th]
	\centering
	\begin{subfigure}[]{.47\linewidth}
		\includegraphics[width=\linewidth]{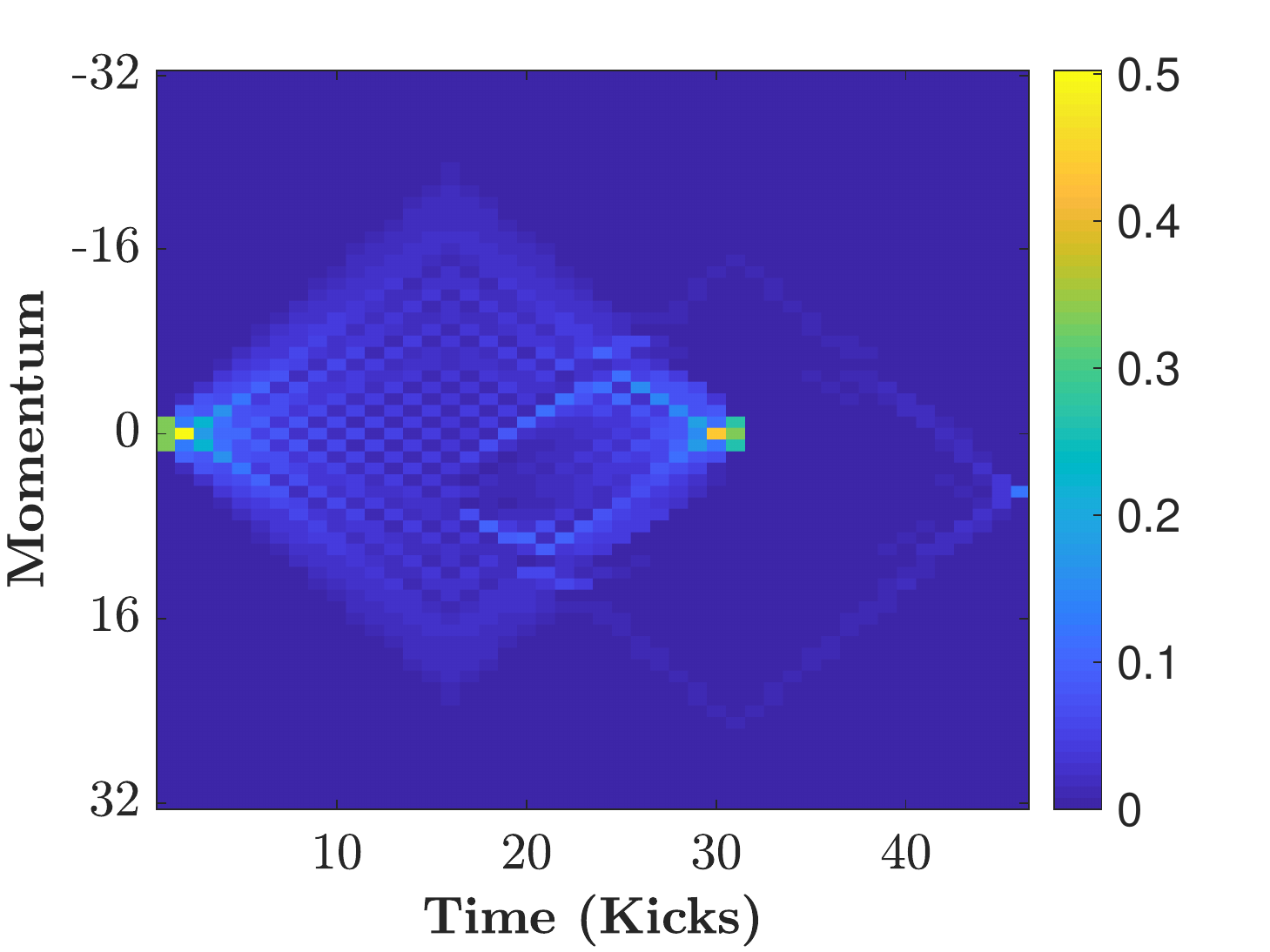}
		\caption{}
		\label{fig:3a}
	\end{subfigure}
	\begin{subfigure}[]{.47\linewidth}
		\includegraphics[width=\linewidth]{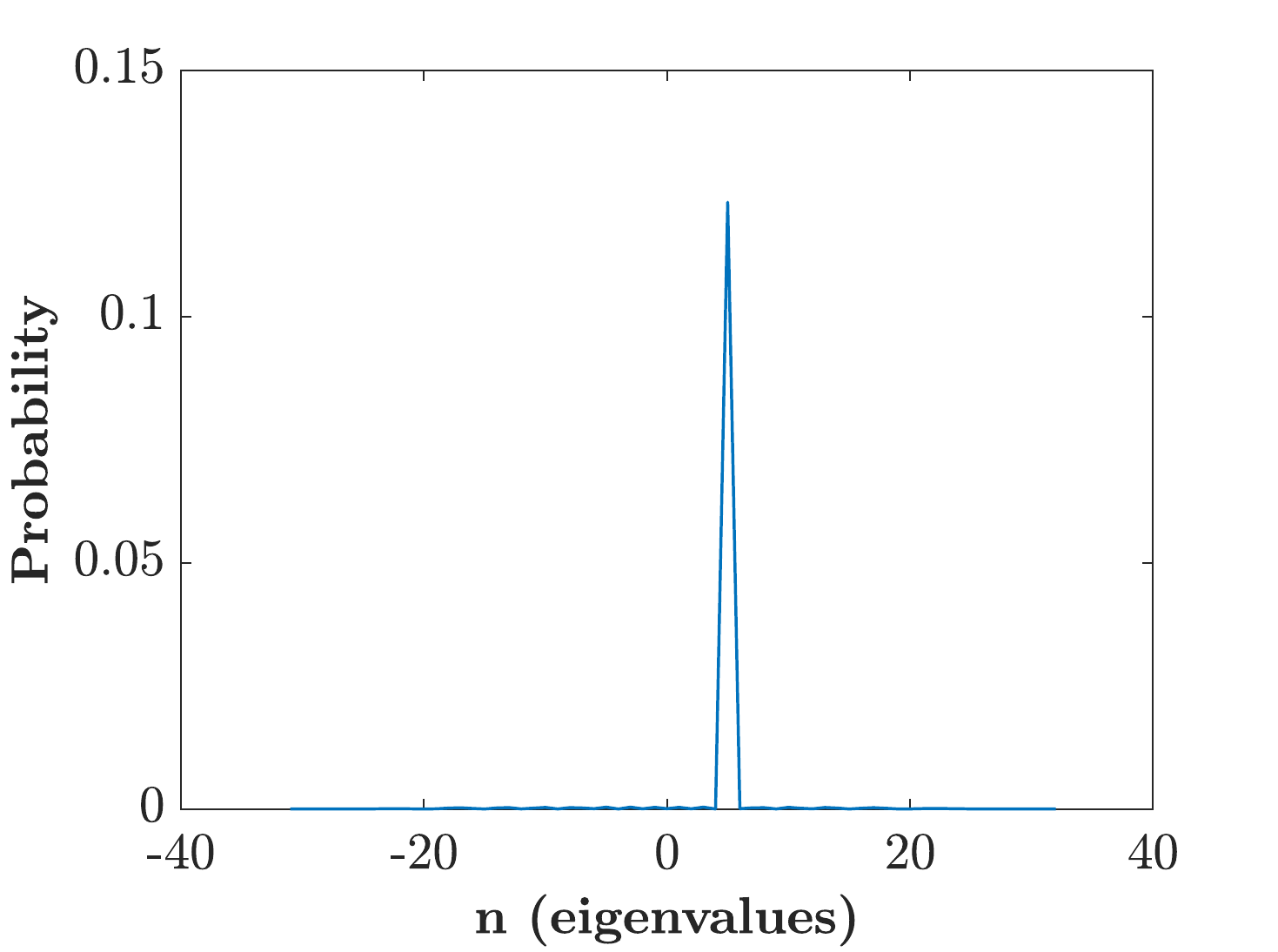}
		\caption{}
		\label{fig:3b}
	\end{subfigure}
	\caption{
	(a) Inversion of the evolution at $t=30$ without considering the central part around $n=0$. (b) The evolution automatically focusses at the target state after returning at $t=45$. The total duration of the protocol is $T=3 \times t = 3 \times 15$ in our example.
	}
		\label{fig:3}
\end{figure}

\begin{enumerate}
\item[\nred{\em (a)}] Suppressing the initial state immediately before measuring the final distribution at $t=30$ leads to the momentum distribution shown in Fig. \ref{fig:2b}. From the latter figure we can find the target state by exploiting the position of the peaks. This is possible because of the asymmetry in the final momentum distribution of Fig. \ref{fig:2b} with respect to zero momentum and because of the linear scaling of the motion with time \cite{WGF2003, sadgrove2011pseudoclassical}. The corresponding shift, e.g. of the two peaks at the flanks $n_l$ and $n_r$, is directly related to the position of the target state $n_t=(n_l+n_r)/2$ (valid for target states in the bulk of the distribution, further details are found in Ref. \cite{Thesis2019}).
\item[\nred{\em (b)}] Alternatively, we can let the walk converge to the target state, just by performing further $15$ kicks of the operator $\hat{\mathcal{U}}$ as Fig. \ref{fig:3a} shows. \nred{If the cutting procedure explained above in {\em (i)} is performed at $t=30$, the central initial state of the distribution will be deleted and we can} obtain a momentum distribution composed essentially only of a single peak centred at the wanted state. This is seen in Fig. \ref{fig:3b}.
\end{enumerate}

Which of the two alternatives is best used in practice is decided by the experiment. Typical signal-to-noise ratios in the measured momentum distributions allow the experimentalist to detect about three orders of magnitude in population difference. This should be sufficient to detect 0.1 (see Fig. \ref{fig:3b}) with respect to the maximum of about 0.5 at the parts which were cut in our example.

\section{Temporal scaling of our search protocol and of our CTQW}
\label{sec:3}

The total time of our protocol is linear in the size of the basis set in which the search takes place. The origin of the linear spread in time is the ballistic motion at QR in the AOKR \cite{sadgrove2011pseudoclassical}. Hence, a QW as ours is always more efficient than a simple classical random (diffusive) walk, whose width would only spread proportionally to the square root of time. 

It can be shown that the optimal time complexity for quantum searching an unsorted database is bounded from below asymptotically by the square root of the number of elements in it, see e.g. \cite{bennett1997strengths}. It should be noted, however, that the dimensionality of the problem heavily influences the efficiency of the algorithms, e.g., for low-dimensional spatial searches, quantum algorithms do not necessarily outperform classical search ones \cite{2000quant.ph..3006B,PhysRevA.78.012310,jeong2013experimental}. Despite no speed-up for searching in a database  with respect to the classical linear search (brute-force searching) for such a simple one-dimensional walk \cite{QA2016}, our proposal is nevertheless interesting for a proof-of-principal realization of searching based on the concrete AOKR experiment.

In the context of classical walks and search, an interesting question is how fast the walk returns to its starting state, the so-called recurrence probability as a function of time. In our case of a quantum walker, the wave function always keeps a final overlap with the initial state. This defines the survival probability, or dynamical fidelity \cite{PROT2016}, for our CTQW. For QR dynamics of the AOKR, the latter is known analytically, and scales as a power law proportional to $1/t$, see e.g. \cite{WB2006, sadgrove2011pseudoclassical, PROT2016}. This power law decay means that the infinite sum of probabilities to find the walker at the origin at any time $\sum_t p_0(t)$ diverges. It follows that the P\'olya number, i.e. the probability to measure the walker at the origin at some stage of the walk \cite{PhysRevLett.100.020501}, is equal to one and our CTQW is therefore called recurrent.

Another measure that is typically investigated in this context is the hitting time, the average time taken by a classical or quantum walk to hit a target node or state. Under the right conditions quantum walks can have much faster hitting times \cite{kempe2005discrete, childs2002example}. One way of defining a quantum hitting time is the time at which the population of the marked state exceeds a chosen threshold (one-shot quantum hitting time). Whatever definition we use, such a hitting time will also scale linearly in the number of steps for our CTQW because of what has been said above in this section.

\section{Conclusions and Outlook}
\label{sec:concl}

We have exploited the analogy of the AOKR with a continuous-time quantum walk to implement a quantum search protocol. In particular, we first created an "optimized", i.e. as flat as possible in some predefined window, probability distribution by forward evolution. This distribution then operates as an initial state for the proper quantum search. After marking a specific target state in the walker's basis, we propagate backwards such that we arrive at an interference pattern created by the walk of the marked state and by the one of the rest. The marked state can be extracted from the information hidden in the final momentum distributions. The best way to do this is to perform another walk with forward and backward evolution of the marked state, cutting away the contribution of the otherwise dominating state around zero momentum (the original initial state).
Then the resulting distribution will be automatically focussed onto the target state to be searched, with a relative probability weight of about ten percent for our time scales of a few tens of steps. Since QKR experiments were performed routinely up to even a few hundred kicks, see e.g. \cite{Schlunk2003,Jones2007,Summy2006,Summy2013,Garreau2015}, we expect that our protocol can be readily realized with a state-of-the-art experiment.

Here we have addressed a one-dimensional quantum walk, therefore a natural extension to be studied would be a multi-dimensional walk using also different topologies. Moreover, the experimental realization of a discrete-time quantum walk with a BEC has already investigated a number of decoherence effects \cite{PRA2019}, affecting the superposition of states and destroying useful information. An extension of such studies to the here proposed CTQW, possibly also including atom-atom interactions, which would act globally in momentum space for a BEC \cite{AW2017}, would be of interest for testing the robustness of our quantum search protocol.

\section*{References}

\bibliographystyle{iopart-num}


\providecommand{\newblock}{}

%

\end{document}